\documentclass[reprint, double column, superscriptaddress,prl, showkeys]{revtex4-1}

\usepackage{float}
\usepackage{graphicx,epsfig}
\usepackage{amssymb}
\usepackage{amsmath}
\usepackage{bm}
\usepackage{braket}

\usepackage{makecell}
\usepackage{textcomp}
\usepackage{color}
\usepackage{pgffor}
\usepackage{pdfpages}

\usepackage{soul}

\usepackage[bookmarksnumbered,bookmarksopen]{hyperref}
\hypersetup{
   colorlinks=true,       
}

\usepackage{BOONDOX-cal}

\makeatletter

\AtBeginDocument{\let\LS@rot\@undefined}
\makeatother
\newif\ifarXiv
\arXivtrue

\usepackage{soul}

\begin{document}
\setcounter{page}{1}

\title[]{Interlayer-correlated fractional quantum Hall state in a trilayer electron system}
\author{Chengyu \surname{Wang}}
\author{C. T. \surname{Tai}}
\author{N. \surname{Toemtrisna}}
\author{A. \surname{Gupta}}
\author{L. N. \surname{Pfeiffer}}
\author{K. W. \surname{Baldwin}}
\author{M. \surname{Shayegan}}
\affiliation{Department of Electrical and Computer Engineering, Princeton University, Princeton, New Jersey 08544, USA}

\date{\today}

\begin{abstract}
{In multilayer quantum Hall systems, when the layer separation is sufficiently small, the interplay between intralayer and interlayer Coulomb interactions can lead to exotic, multi-component, many-body states. A particular example is the even-denominator fractional quantum Hall state (FQHS) at total filling factor $\nu=$ 1/2 in bilayer systems with negligible interlayer tunneling. This state can be understood as a generalized Laughlin state described by the Halperin-Laughlin $\Psi_{331}$ wavefunction. While such correlated states have been extensively explored in bilayers, little is known about their counterparts in systems with more than two layers. Here, we investigate a trilayer two-dimensional electron system confined to ultrahigh-quality GaAs triple quantum wells. We observe an exotic FQHS at total filling factor $\nu=$ 5/7, evinced by a deep longitudinal resistance minimum and a quantized Hall plateau, when the side layers have a density larger than the middle layer and $d/l_B \simeq 2.2$ ($d$ is the interlayer distance and $l_B$ the magnetic length). This state is naturally interpreted as the long-predicted trilayer, generalized Laughlin ($\Psi_{33311}$) state, characterized by $\nu=1/3$-like intralayer correlation within each layer and $\nu=1$-like interlayer correlation between neighboring layers. Our observation establishes a new member of the Halperin–Laughlin many-body states that extends interlayer coherence to three coupled layers. }
\end{abstract}

\maketitle

Under strong perpendicular magnetic fields and at low temperatures, electrons in a two-dimensional electron system (2DES) are confined to quantized Landau levels, quenching kinetic and thermal energies and allowing Coulomb interaction to dominate. In this interaction-driven regime, a plethora of correlated quantum phases emerges, most notably the fractional quantum Hall states (FQHSs)—incompressible electron liquids characterized by fractional charge, anyonic statistics, and nontrivial topological order~\cite{Tsui.PRL.1982, Laughlin.PRL.1983, Jain.PRL.1989, Jain.Book.2007, Halperin.Book.2020}.

Additional internal degrees of freedom, such as spin, valley, layer, and electric subband, further enrich FQHS physics. Among these, the layer degree of freedom is particularly compelling: when two layers are brought sufficiently close, the competition between intralayer and interlayer Coulomb interactions gives rise to correlated phases without single-layer analogues. These include interlayer-correlated, two-component FQHSs~\cite{Eisenstein.PRL.1992, Liu.NatPhys.2019, Li.NatPhys.2019, Zhang.Nature.2025}, exciton condensate superfluids at total filling factor $\nu_{\text{total}}=1$~\cite{Murphy.PRL.1994, Eisenstein.Nature.2004, Kellogg.PRL.2004, Tutuc.PRL.2004, Wiersma.PRL.2004, Nandi.Nature.2012, Eisenstein.ARCMP.2014, Liu.NatPhys.2017, Li.NatPhys.2017}, and bilayer Wigner crystal phases~\cite{Manoharan.PRL.1996, Hatke.PRB.2017}. Intriguingly, theory~\cite{Nomura.JPSJ.2004, Peterson.PRB.2010, Zhu.PRB.2016} and experiment~\cite{Suen.PRL.1992, Mueed.PRL.2015, Mueed.PRL.2016, Singh.NatPhys.2024, Singh.PRL.2025} suggest that a large interlayer tunneling can drive a two-component to one-component phase transition in the even-denominator FQHS at $\nu=1/2$, opening up new pathways for realizing non-Abelian anyons for topological quantum computing~\cite{Nayak.RMP.2008}. Although FQHSs have been extensively studied in bilayer systems using various material platforms such as GaAs double and wide quantum wells and double-layer graphene, they are largely unexplored in a trilayer setup~\cite{Jo.PRB.1992, Shukla.PRL.1998, Gusev.PRB.2009, Wiedmann.JPCS.2011}.

\begin{figure*}[t]
  \begin{center}
    \psfig{file=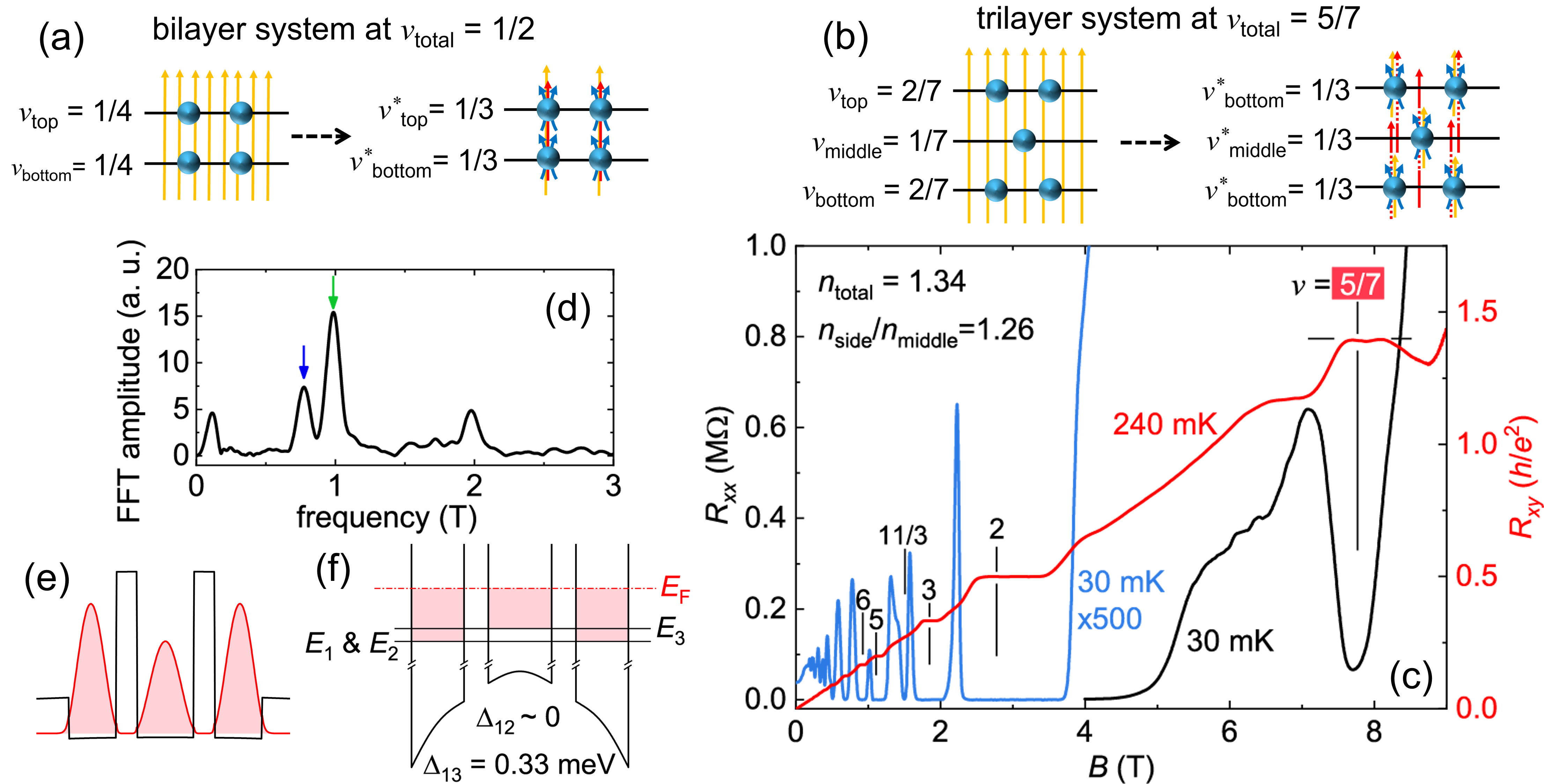, width=0.95 \textwidth}
  \end{center}
  \caption{\label{main}
    {\bf Interlayer-correlated FQHS in a trilayer 2DES.} (a, b) Schematic descriptions of the interlayer-correlated FQHSs in the weakly-interacting CF picture for: (a) bilayer 2DESs at $\nu_{\text{total}}=1/2$, and (b) trilayer 2DESs at $\nu_{\text{total}}=5/7$. The blue spheres represent electrons, and the yellow vertical arrows the magnetic field flux quanta. Each electron captures two flux quanta from its own layer (blue curved arrows) and one flux quantum from its neighboring layer(s) (red arrows), forming a multi-component ($2+1$)-flux CF. $\nu_{\text{top}}$, $\nu_{\text{middle}}$, and $\nu_{\text{bottom}}$ are the actual layer fillings and $\nu^*_{\text{top}}$, $\nu^*_{\text{middle}}$, and $\nu^*_{\text{bottom}}$ are the effective layer fillings that reflect the intralayer correlation. (c) $R_{xx}$ and $R_{xy}$ vs $B$ data for our triple-QW sample measured at 30 mK and 240 mK, respectively. The trilayer 2DES has a total density of 1.34$\times10^{11}$ cm$^{-2}$ and a symmetric charge distribution. At zero and low $B$, the densities in the top, middle, and bottom QWs have a ratio of $1.26:1:1.26$. A pronounced FQHS is observed at $\nu_{\text{total}}=5/7$ evinced by a deep $R_{xx}$ minimum and a quantized Hall plateau. (d) FFT of low-$B$ ($\sim$0.04 to $\sim$0.6 T) SdH oscillations. The blue and green arrows mark the peaks corresponding to the SdH oscillations of the middle and side QWs, respectively. (e) Self-consistently calculated charge distribution (red) and potential energy (black) for our triple-QW sample. (f) Expanded view of the potential energy, showing the subband energy levels and the Fermi energy.
    }
  \label{fig:main}
\end{figure*}

Multilayer FQHSs can be divided into two categories according to the presence or absence of interlayer correlation. The latter depends on the ratio of intralayer and interlayer Coulomb interaction $d/l_B$, where $d$ is the interlayer distance and $l_{B}$ the magnetic length. For $d \gg l_B$, intralayer interactions dominate and the layers host independent FQHSs. In bilayer systems, such FQHSs are observed at total filling factor $\nu_{\text{total}}$ with even numerators or non-prime denominators, e.g. at $2/3$ ($\nu=1/3$ for each layer) and $11/15$ ($\nu=1/3$ for one layer and $\nu=2/5$ for the other layer)~\cite{Monoharan.PRL.1997}. When $d\sim l_B$, a different type of multilayer FQHS, one with interlayer correlation emerges~\cite{Halperin.HPA.1983, MacDonald.SurfSci.1990, Eisenstein.PRL.1992, Scarola.PRB.2001, Liu.NatPhys.2019, Li.NatPhys.2019, Zhang.Nature.2025}. The simplest interlayer-correlated FQHS is the so-called $\Psi_{331}$ state in a bilayer system at $\nu_{\text{total}}=1/2$~\cite{Halperin.HPA.1983, MacDonald.SurfSci.1990, Eisenstein.PRL.1992, Scarola.PRB.2001}; see Fig.~\ref{fig:main}(a). It can be described by the generalized Laughlin wave function, first proposed by Halperin~\cite{Halperin.HPA.1983}:
\begin{equation}
   \Psi_{331} =
\prod_{i<j}(z_i - z_j)^3
\prod_{i<j}(w_i - w_j)^3
\prod_{i,j}(z_i - w_j)^1.
\end{equation}
Here, $z_i$ and $z_j$ are the complex coordinates of electrons in the upper layer, and $w_i$and $w_j$ are those in the lower layer. The $\Psi_{331}$ state is stabilized by the strong $\nu=1/3$-like intralayer correlation and relatively weaker $\nu=1$-like interlayer correlation, and can be well explained by the two-component, ($2+1$)-flux composite fermion (CF) model~\cite{Jain.Book.2007, Scarola.PRB.2001}; see Fig.~\ref{fig:main}(a) for a schematic description of the $\Psi_{331}$ state in the CF model.

One would expect that similar generalized Laughlin states with interlayer correlation should also exist in systems with three or more layers~\cite{MacDonald.SurfSci.1990}. In a trilayer system, the interaction between the top and bottom layers is rather weak because of the relatively large interlayer distance and the screening provided by the middle layer. By considering strong intralayer and nearest-neighbor interlayer Coulomb interactions, the strongest interlayer-correlated, generalized Laughlin state in a trilayer system is predicted to occur at $\nu_{\text{total}}=5/7$ when the system has layer fillings 2/7, 1/7, and 2/7 for the top, middle, and bottom layers, respectively~\cite{MacDonald.SurfSci.1990}. Such a state is described by the $\Psi_{33311}$ wave function:
\begin{multline}
\Psi_{33311} =
\prod_{i<j}(z_i - z_j)^3
\prod_{i<j}(w_i - w_j)^3
\prod_{i<j}(v_i - v_j)^3 \\
\times
\prod_{i,j}(z_i - w_j)^1
\prod_{i,j}(v_i - w_j)^1.
\end{multline}
where $z_i$, $w_i$, and $v_i$ are the complex coordinates of electrons in the top, middle, and bottom layers, respectively. It is characterized by $\nu=1/3$-like intralayer interaction in each layer, and $\nu=1$-like interlayer interaction between neighboring layers; see Fig.~\ref{fig:main}(b) for a CF description. Experimentally, trilayer quantum Hall physics has been explored in GaAs triple quantum wells (QWs)~\cite{Jo.PRB.1992, Shukla.PRL.1998, Gusev.PRB.2009, Wiedmann.JPCS.2011}. An unusually deep longitudinal resistance ($R_{xx}$) minimum was indeed observed at $\nu_{\text{total}}=5/7$~\cite{Jo.PRB.1992}. However, neither a Hall plateau nor an energy gap were reported, leaving it an open question whether or not the 5/7 state is a true FQHS.

Here we report the observation of an interlayer-correlated trilayer FQHS in a 2DES confined to GaAs triple QWs. As highlighted in Fig.~\ref{fig:main}(c), this FQHS is observed at $\nu_{\text{total}}=5/7$, signaled by a very deep $R_{xx}$ minimum and a Hall resistance ($R_{xy}$) plateau quantized at $7h/5e^2$. We find that the $\nu_{\text{total}}=5/7$ FQHS is stabilized when: (\textit{i}) the charge distribution is symmetric, and (\textit{ii}) the side layers' density is larger than the middle layer density, qualitatively in agreement with the predicted $\Psi_{33311}$ state.

We studied an ultra-high-quality, triple-layer 2DES confined to three GaAs QWs grown on GaAs (001) substrates by molecular beam epitaxy. The sample was grown following the optimization of the growth chamber vacuum integrity and the purity of the source materials~\cite{Chung.NatMater.2021}. The widths of the middle and side QWs are 158 and 132 $\text{\r{A}}$, respectively. The middle QW is designed to be wider than the side QWs to enhance electron population in the middle QW. The QWs are separated by 55-$\text{\r{A}}$-wide pure AlAs barriers. The interlayer tunneling between the neighboring layers is negligible. We performed our experiments on a $4\times 4$ mm$^2$, van der Pauw geometry sample cleaved from a 2-inch GaAs wafer. Ohmic contacts to all three layers were formed using alloyed In:Sn at the four corners and side midpoints. The sample was fitted with a Ti/Au front gate and an In back gate, allowing for \textit{in-situ} tuning of 2D electron density and charge distribution; see SM for details \cite{SM}. The total as-grown density of the trilayer electron system is $1.98\times10^{11}$ cm$^{-2}$, and its mobility (measured at 0.3 K) is $\simeq 8.2\times10^5$ cm$^2$/Vs, about three times larger than in previous studies \cite{Jo.PRB.1992}. The sample was cooled in a dilution refrigerator with a base temperature of $\simeq 30$ mK. We measured $R_{xx}$ and $R_{xy}$ using the conventional, low-frequency ($\sim17$ Hz), lock-in amplifier technique.

Figure~\ref{fig:main}(c) presents $R_{xx}$ and $R_{xy}$ vs $B$ traces of our triple-QW sample when the sample is carefully tuned to a total density of $n_{\text{total}}=1.34$ (in units of $10^{11}$ cm$^{-2}$, which we use throughout the Letter) with symmetric charge distribution~\cite{SM}. The data demonstrate a FQHS at $\nu_{\text{total}}=5/7$: We observe a deep $R_{xx}$ minimum at $\nu_{\text{total}}=5/7$ accompanied by an $R_{xy}$ plateau quantized at $7h/5e^2$. The $\nu_{\text{total}}=5/7$ FQHS is flanked by highly-resistive phases with $R_{xx}$ values reaching several hundred k$\Omega$. The data are reminiscent of the insulating phases observed on the flanks of the $\nu=1/5$ FQHS in single-layer GaAs 2DESs~\cite{Jiang.PRL.1990, Goldman.PRL.1990}, as well as those on the flanks of the $\nu=1/2$ FQHS in bilayer GaAs 2DESs~\cite{Eisenstein.PRL.1992, Manoharan.PRL.1996}. These insulating phases are believed to be single-layer~\cite{Jiang.PRL.1990, Goldman.PRL.1990, Deng.PRL.2016} and bilayer Wigner crystal states~\cite{Manoharan.PRL.1996, Hatke.PRB.2017, Zuo.PRB.2020, Faugno.PRB.2018}. The insulating phases we observe suggest the emergence of trilayer Wigner crystal states reentrant around the $\nu_{\text{total}}=5/7$ FQHS. It is worth noting that measuring Hall resistance in a highly-insulating regime is extremely challenging because of the nonuniform current distribution and mixing of the very large $R_{xx}$ signal into the $R_{xy}$ measurements. We minimize such effects by increasing the temperature to 240 mK to reduce $R_{xx}$ and antisymmetrizing $R_{xy}$ measured for opposite polarities of $B$; see SM for details \cite{SM}. At lower $B$, numerous integer quantum Hall states (at $\nu_{\text{total}}=2$, 3, 5, 6, ...) and a FQHS (at $\nu_{\text{total}}=11/3$) are identified.

In Fig.~\ref{fig:main}(d), we present fast Fourier transform (FFT) of the low-$B$ Shubnikov-de Haas (SdH) oscillations data. From the evolution of the FFT peaks as a function of gate voltages, as shown in Figs. S1 and S2 of SM \cite{SM}, we determine that the two pronounced peaks in the spectrum, marked by blue and green arrows, correspond to the SdH oscillations of the middle QW and side QWs, respectively. From the FFT peak positions, we deduce the low-$B$, middle-layer density ($n_{\text{middle}}=0.38$) and side-layer density ($n_{\text{side}}=0.48$); see Figs.~\ref{fig:main}(e,f) for charge distribution, conduction band energy, subband energy levels, and Fermi energy obtained from self-consistent Hartree calculations at zero $B$.

\begin{figure}[t]
  \begin{center}
    \psfig{file=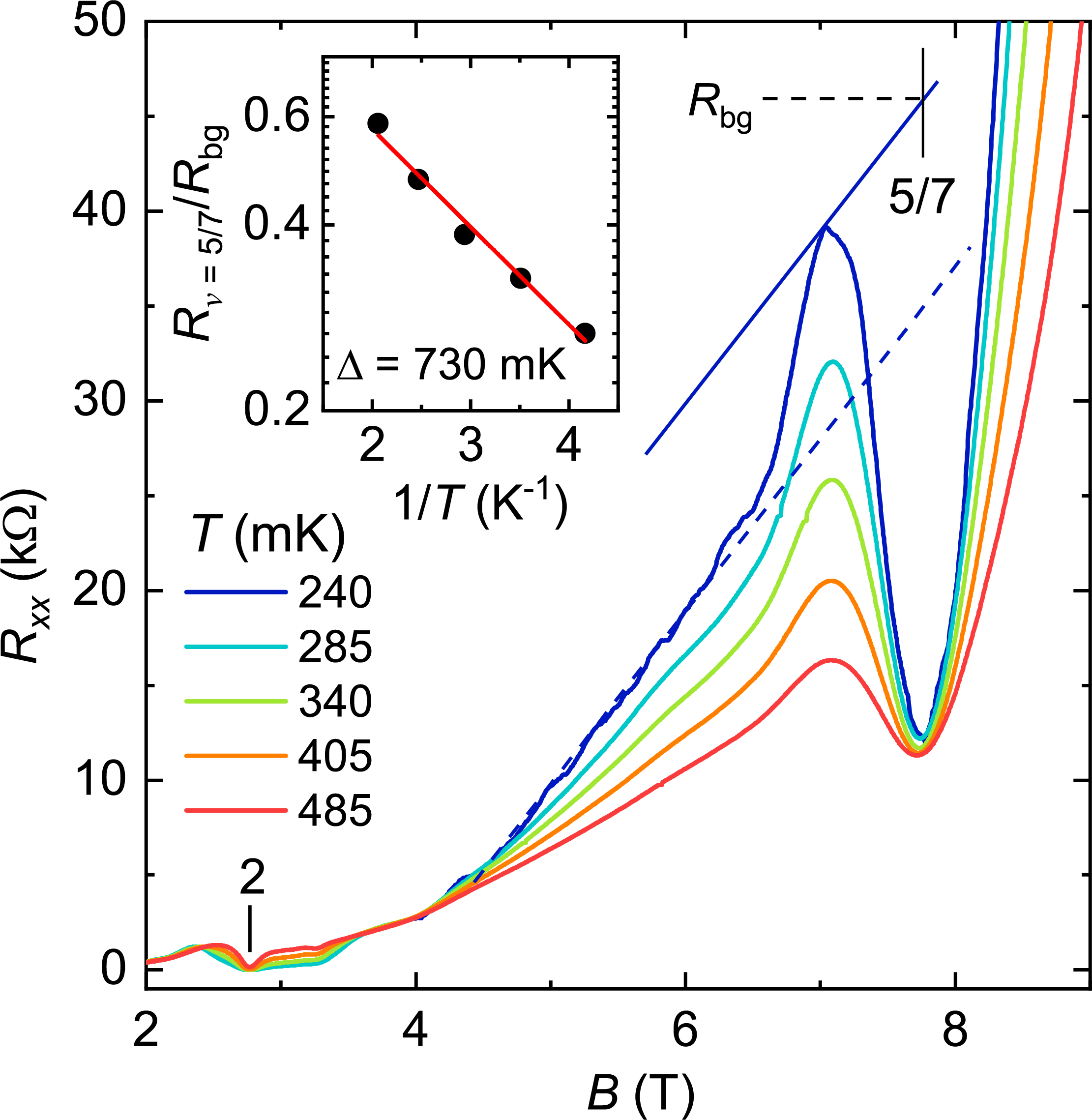, width=0.47 \textwidth}
  \end{center}
  \caption{\label{temp} 
   {\bf Temperature dependence data.} $R_{xx}$ vs $B$ traces at $n_{\text{total}}=1.34\times10^{11}$ cm$^{-2}$, measured at different temperatures. The dashed and solid lines illustrate, using the 240 mK data as an example, the method used to determine the background resistance $R_{\mathrm{bg}}$ at each temperature \cite{Mallett.PRB.1988}. Inset shows the Arrhenius plot of $R_{\nu=5/7}/R_{\mathrm{bg}}$, where $R_{\nu=5/7}$ is the $R_{xx}$ value at $\nu_{\text{total}}=5/7$. Slope of the linear fit provides an estimate of the energy gap for the $5/7$ FQHS, $\Delta\simeq730$ mK.}
  \label{fig:temp}
\end{figure}

The temperature dependence of the FQHS at $\nu_{\text{total}}=5/7$ is presented in Fig.~\ref{fig:temp}. The transport energy gap ($\Delta$) of a FQHS is typically deduced from the relation $R_{xx}\propto e^{-\Delta/2k_BT}$. For our electron system, however, with decreasing temperature, instead of approaching zero, $R_{xx}$ at $\nu_{\text{total}}=$ 5/7 increases because of the temperature-dependent $R_{xx}$ background. A useful method to estimate the energy gap of a FQHS riding on a temperature-dependent resistance background is to extract a so-called pseudogap from the temperature dependence of the relative depth of the $R_{xx}$ minimum~\cite{Mallett.PRB.1988}. In Fig.~\ref{fig:temp} inset, we present Arrhenius plots of $R_{\nu=5/7}/R_{\mathrm{bg}}$ vs $1/T$, where $R_{\mathrm{bg}}$ is the background resistance. Slope of the linear fit gives an estimate of the energy gap for the $5/7$ FQHS, $\Delta\simeq730$ mK.

\begin{figure*}[t]
  \begin{center}
    \psfig{file=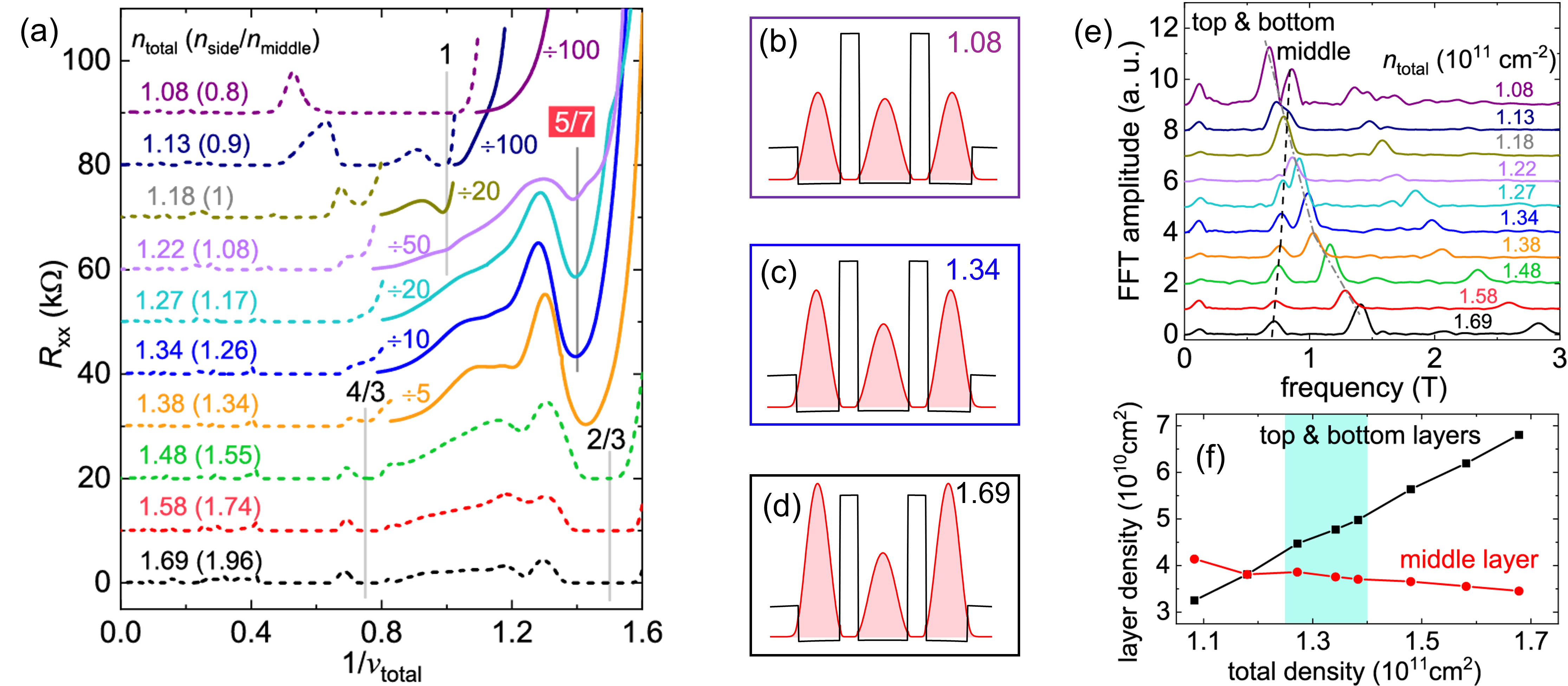, width=0.95 \textwidth}
  \end{center}
  \caption{\label{sym} 
   {\bf Evolution of FQHSs with symmetric gating.} (a) $R_{xx}$ vs $1/\nu_{\mathrm{total}}$ traces measured at different total densities, while the front and back gates are used to ensure that the top and bottom QWs have equal densities. Each trace is vertically shifted by 10 k$\Omega$ for clarity. (b-d) Charge distributions and potentials for $n_{\text{total}}=$ 1.08, 1.34, and 1.69, calculated self-consistently at zero magnetic field. (e) FFT spectra of low-$B$ SdH oscillations at different $n_{\text{total}}$. From the frequencies of the FFT peaks, we determine the layer densities, as shown in (f). The density range where a $\nu_{\text{total}}=5/7$ FQHS is observed is highlighted.}
  \label{fig:sym}
\end{figure*}

We note that the $\nu_{\text{total}}=5/7$ FQHS is observed when $n_{\text{side}}/n_{\text{middle}}=1.26$ [Fig.~\ref{fig:main}(c)]. This is surprising at first glance, because the theoretically predicted $\Psi_{33311}$ state has layer fillings $\nu_{\text{side}}=2/7$ and $\nu_{\text{middle}}=1/7$, and should be stabilized when $n_{\text{side}}/n_{\text{middle}}=2$. As we discuss below, the apparent discrepancy stems from the significant difference in $n_{\text{side}}/n_{\text{middle}}$ at low $B$ and high $B$~\cite{Shukla.PRL.1998, Shukla.reply}.

In Fig.~\ref{fig:sym}(a), we present $R_{xx}$ vs $1/\nu$ traces measured at different $n_{\text{total}}$; the corresponding $R_{xx}$ and $R_{xy}$ vs $B$ traces are provided in Figs. S5 and S6 of the SM \cite{SM}. The density is tuned symmetrically using both front and back gates while ensuring the top and bottom QWs have equal densities [see Figs.~\ref{fig:sym}(b-d) for charge distributions]. We determine layer densities from FFT spectra of low-$B$ SdH oscillations shown in Fig.~\ref{fig:sym}(e). As we increase $n_{\text{total}}$, $n_{\text{side}}$ increases because charges are added to the top and bottom QWs, while $n_{\text{middle}}$ slightly decreases because of the spontaneous, exchange-induced, interlayer charge transfer [Fig.~\ref{fig:sym}(f)]~\cite{Ying.PRB.1995, Papadakis.PRB.1997, Eisenstein.PRB.1994, Lay.PRB.1995}.

In Fig.~\ref{fig:sym}(a), the strongest $\nu_{\text{total}}=5/7$ FQHS is observed at $n_{\text{total}}=1.34$ ($n_{\text{side}}/n_{\text{middle}}=1.26$). As we decrease $n_{\text{total}}$ (so that $n_{\text{side}}/n_{\text{middle}}$ decreases), the $R_{xx}$ minimum at $\nu_{\text{total}}=5/7$ becomes weaker, while the $R_{xx}$ background increases and becomes immeasurably large, as the $5/7$ FQHS gives way to an insulating phase. The evolution with increasing $n_{\text{total}}$ is remarkably different. As we increase $n_{\text{total}}$, the $R_{xx}$ minimum at $\nu_{\text{total}}=5/7$ gradually moves towards lower $\nu_{\text{total}}$, and eventually becomes a broad minimum which approaches zero at $\nu_{\text{total}}=2/3$. Meanwhile, the insulating phases at low fillings ($\nu_{\text{total}}\lesssim1$) disappear. The trace at $n_{\text{total}}=1.69$ exhibits a bilayer behavior at $\nu_{\text{total}}<2$: FQHSs are observed at even-numerator $\nu_{\text{total}}=2/3$, 4/5, 4/3, and 6/5 (see Fig. S5 in SM for both $R_{xx}$ and $R_{xy}$ data~\cite{SM}), consistent with two decoupled FQHSs in the top and bottom layers with layer fillings $\nu_{\text{middle}}=0$ and $\nu_{\text{side}}=$ 1/3, 2/5, 2/3, and 3/5. This suggests that the electron system is making a trilayer to bilayer transition when $n_{\text{side}}/n_{\text{middle}}$ is increased from 1.26 to 1.96, although the middle QW still has significant amount of charge at low $B$. A similar trilayer to bilayer transition at $\nu_{\text{total}}<2$ was also reported in GaAs triple QWs with a larger interlayer tunneling~\cite{Shukla.PRL.1998, Shukla.reply}. Such a transition suggests that $n_{\text{side}}/n_{\text{middle}}$ at high $B$ ($\nu_{\text{total}}<2$) is significantly larger than its value at $B\simeq0$. This is the result of the combined effect of Landau quantization, and intra- and interlayer exchange energies~\cite{Hanna.PRB.1996}. Therefore, it is likely that at $n_{\text{total}}=1.34$, $n_{\text{side}}/n_{\text{middle}}$ is $\simeq$ 2 at $\nu_{\text{total}}=5/7$, much larger than its value (1.26) measured at low $B$, and consistent with the theoretical $\Psi_{33311}$ state~\cite{MacDonald.SurfSci.1990}.

\begin{figure}[t]
  \begin{center}
    \psfig{file=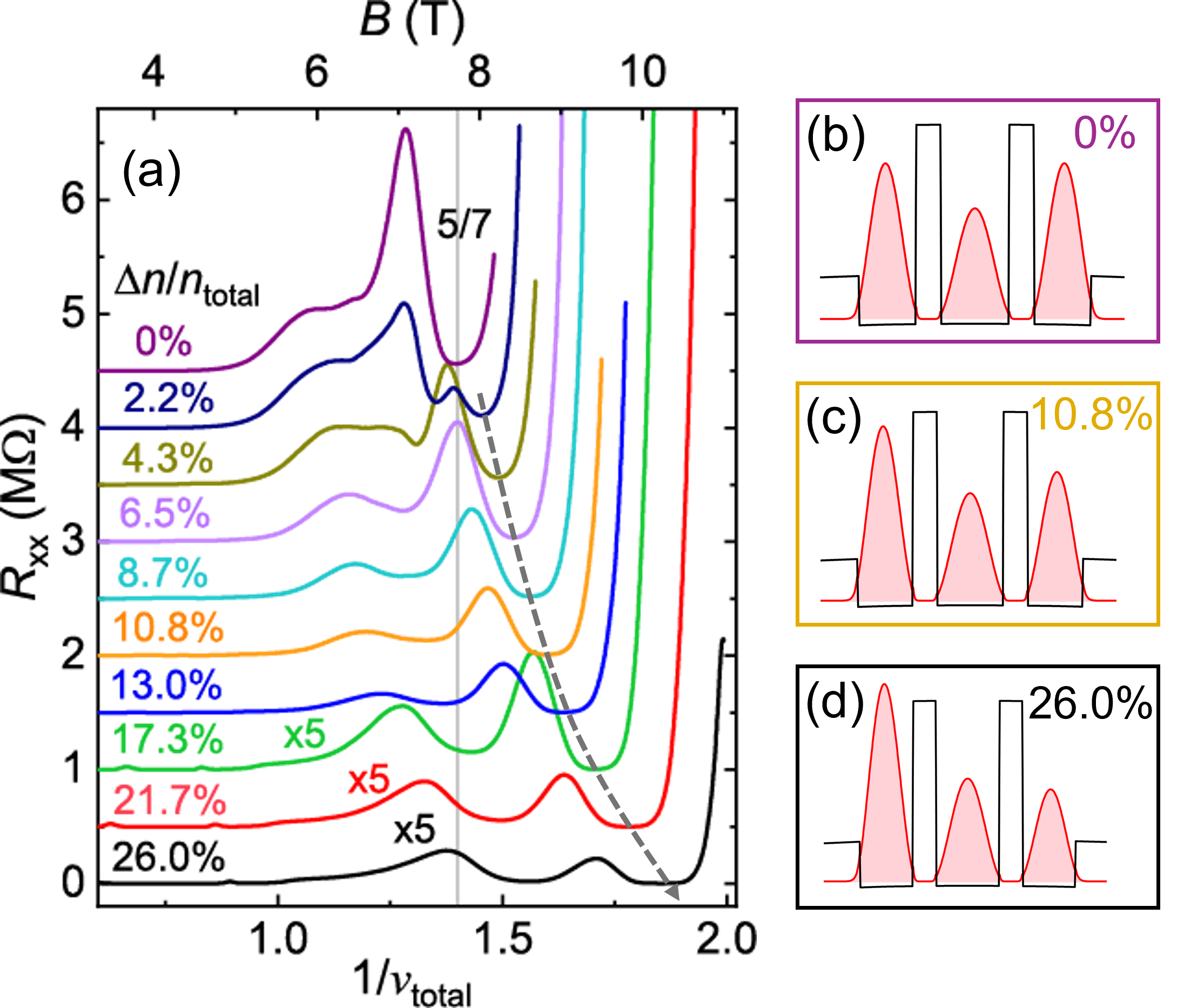, width=0.43 \textwidth}
  \end{center}
  \caption{\label{imbalance} 
   {\bf Effect of layer imbalance.} (a) $R_{xx}$ vs $B$ traces measured at fixed $n_{\text{total}}=1.34$, but different $\Delta n/n_{\text{total}}$ (shown on the left for each trace), where $\Delta n$ is the density difference between the top and bottom layers. The traces for $\Delta n/n_{\text{total}}=17.3\%$, $21.7\%$, and $26.0\%$ are multiplied by 5 to show the features clearly. (b-d) Self-consistently calculated charge distributions at $B=0$, for $\Delta n/n_{\text{total}}=0$, $10.8\%$, and $26.0\%$.}
  \label{fig:imbalance}
\end{figure}

Next we investigate the fate of the $5/7$ FQHS with charge distribution asymmetry. We increase $n_{\text{bottom}}$ by $\Delta n/2$ using the back gate and decrease $n_{\text{top}}$ by the same amount using the front gate, so that $n_{\text{total}}$ is kept fixed at 1.34; see SM Figs. S1 and S2 for FFT spectra of low-$B$ SdH oscillations and the extracted layer densities~\cite{SM}. Figure~\ref{fig:imbalance} shows $R_{xx}$ vs $B$ traces measured at $n_{\text{total}}=1.34$ and $\Delta n/n_{\text{total}}$ ranging from 0 to $26.0\%$; the corresponding $R_{xy}$ traces and expanded views of the $R_{xx}$ traces are provided in Figs. S7-S9 of the SM \cite{SM}. We note that a small imbalance ($\Delta n/n_{\text{total}}=2.2\%$) destabilizes the $5/7$ FQHS, and leads to two minima in $R_{xx}$ near $\nu_{\text{total}}=5/7$. With increasing $\Delta n/n_{\text{total}}$, the $R_{xx}$ minimum on the high-$B$ side of $\nu_{\text{total}}=5/7$ becomes deeper, and gradually moves towards higher $B$. At the $B$ position where this $R_{xx}$ minimum is seen, we observe an $R_{xy}$ plateau quantized at $3h/e^2$ when $\Delta n/n_{\text{total}}\geq17.3\%$; see SM Figs. S7 and S9~\cite{SM}. The emergence of a quantized $3h/e^2$ plateau suggests that at large imbalance, the bottom layer which has the highest density becomes dominant and forms a single-layer $\nu=1/3$ FQHS, and interlayer correlation is lost.

The evolutions of the $\nu_{\text{total}}=5/7$ FQHS under symmetric and imbalance gating are qualitatively consistent with the theoretically proposed $\Psi_{33311}$ state: (\textit{i}) The $5/7$ FQHS is observed when $n_{top}=n_{bottom}>n_{middle}$ and only in a limited range of $n_{side}/n_{middle}$ before the system makes a trilayer-to-bilayer transition. (\textit{ii}) A very small charge distribution asymmetry ($2.2\%$) destabilizes the $5/7$ FQHS. (\textit{iii}) In GaAs double QWs, the FQHS at $\nu=1/2$, which is believed to be the $\Psi_{331}$ state, was reported in samples with $d/l_B<3$~\cite{Eisenstein.PRL.1992}. This is comparable to the $d/l_B\simeq2.2$ for the $5/7$ FQHS we observe in our triple-QW sample. 

It is worth remarking that another possible origin for the $5/7$ FQHS in our trilayer electron system is a layer-decoupled FQHS with $\nu_{\text{side}}=2/7$ and $\nu_{\text{middle}}=1/7$. This is extremely unlikely. Although FQHSs have been observed at $\nu=2/7$ and 1/7 in single-layer GaAs 2DESs~\cite{Chung.PRL.2022, Wang.PRL.2025}, they are extremely weak, much weaker than those observed at $\nu=2/3$ and 1/3. If the three layers were decoupled, a $\nu_{\text{total}}=5/3$ FQHS with $\nu_{\text{side}}=2/3$ and $\nu_{\text{middle}}=1/3$ should be stronger than the layer-decoupled $\nu_{\text{total}}=5/7$ FQHS. However, we do not observe a FQHS at $\nu_{\text{total}}=5/3$, strongly suggesting that the $5/7$ FQHS in our sample has interlayer correlation.

Our observations establish a correlated FQHS with genuine trilayer interlayer correlations, extending the Halperin–Laughlin many-body states beyond bilayers. Together with the emergence of insulating phases on its flanks, these results reveal a rich interplay between liquid and crystalline order in multilayer quantum Hall systems. More broadly, trilayer and higher-layer platforms offer a fertile setting for realizing multicomponent topological states and for exploring qualitatively new collective phenomena that emerge as the number of coupled layers increases~\cite{Qiu.PRB.1989, Naud.PRL.2000, Wang.arXiv.2025}.

\section*{Acknowledgement} 
We acknowledge support by the National Science Foundation (NSF) Grant Nos. DMR 2104771 and DMR 2611783 for measurements, and the Gordon and Betty Moore Foundation’s EPiQS Initiative (Grant No. GBMF9615.01 to L.N.P.) for sample fabrication. Our measurements were partly performed at the National High Magnetic Field Laboratory (NHMFL), which is supported by the NSF Cooperative Agreement No. DMR 2128556, by the State of Florida, and by the DOE. This research was funded in part by QuantEmX travel grants from Institute for Complex Adaptive Matter and the Gordon and Betty Moore Foundation through Grant GBMF9616 to C. W. and C. T. T. We thank A. Bangura, G. Jones, R. Nowell and T. Murphy at NHMFL for technical assistance. We also thank J. K. Jain for illuminating discussions.



\section*{Supplementary material}

# Supplemental Material to "Interlayer-correlated fractional quantum Hall state in a trilayer electron system"

Chengyu Wang,[1] C. T. Tai,[1] N. Toemtrisna,[1] A. Gupta,[1]
L. N. Pfeiffer,[1] K. W. Baldwin,[1] and M. Shayegan[1]

[1]*Department of Electrical and Computer Engineering,*
*Princeton University, Princeton, New Jersey 08544, USA*

## 1. EXTRACTING LAYER DENSITIES FROM FAST FOURIER TRANSFORM ANALYSIS OF SHUBNIKOV-DE HAAS OSCILLATIONS

In order to determine layer densities for our trilayer electron system and find the balanced point with symmetric charge distribution ($n_{\text{bottom}} = n_{\text{top}}$), we measured Shubnikov-de Haas (SdH) oscillations at fixed total density $n_{\text{total}} = 1.34 \times 10^{11}$ cm$^{-2}$ but different $\Delta n$ ($n_{\text{bottom}} - n_{\text{top}}$). This is achieved by increasing the density using the back gate by $\Delta n/2$ and decreasing the density using the front gate by the same amount. Figure S1 presents fast Fourier transform (FFT) spectra of SdH oscillations measured at $n_{\text{total}} = 1.34 \times 10^{11}$ cm$^{-2}$ and $\Delta n/n_{\text{total}}$ ranging from -4.3% to +26.0%. In most of our FFT spectra, we observe three major peaks, corresponding to SdH oscillations of the three layers. The positions of the major FFT peaks are proportional to the electron densities in the three layers. The black,

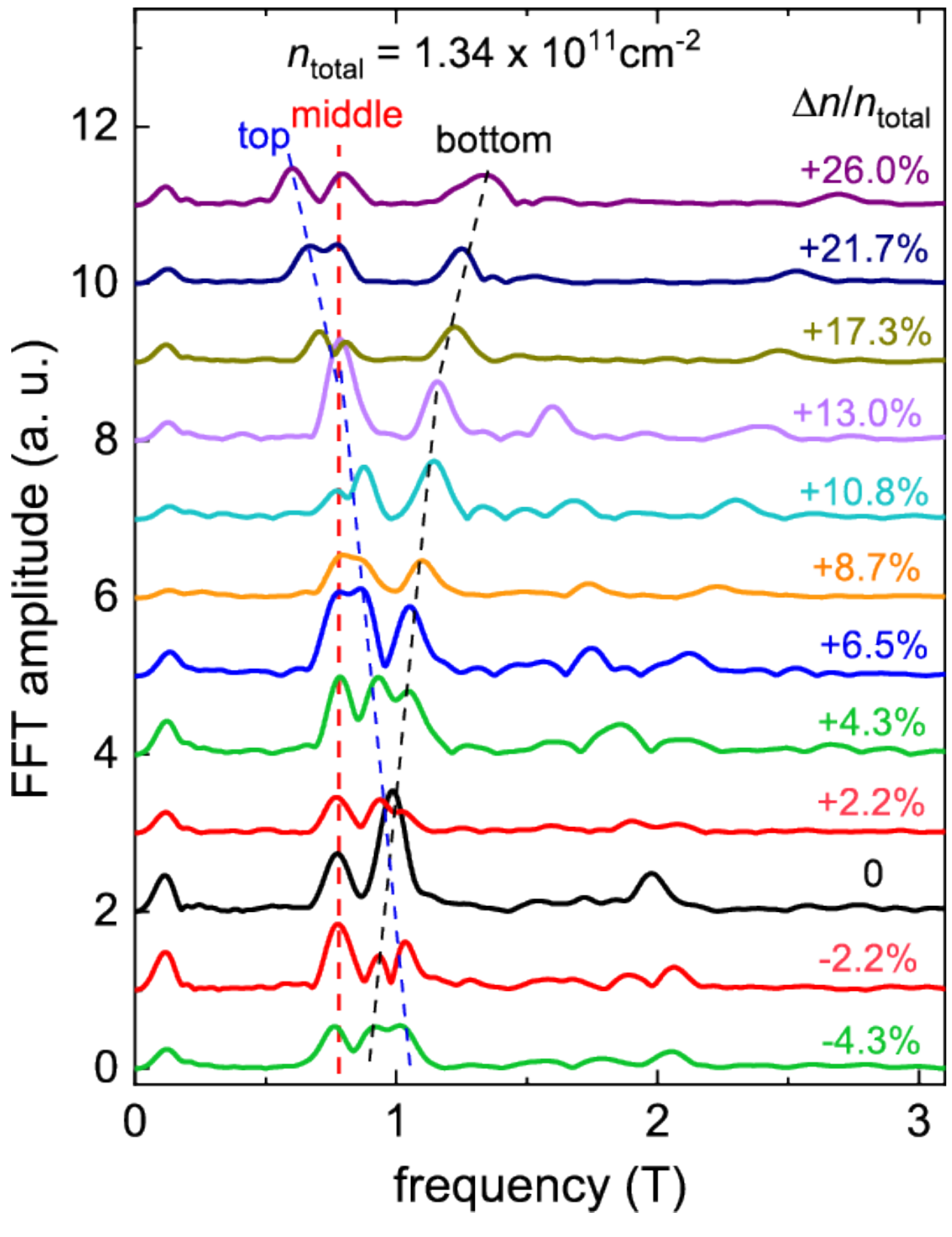


Fig. S1. FFT of low-$B$ (0.04 T to 0.6 T) Shubnikov-de Haas oscillations measured at total density $n_{\text{total}} = 1.34 \times 10^{11}$ cm$^{-2}$ and $\Delta n/n_{\text{total}}$ ranging from -4.3% to +26.0%, where $\Delta n$ is the density difference between bottom and top layers.

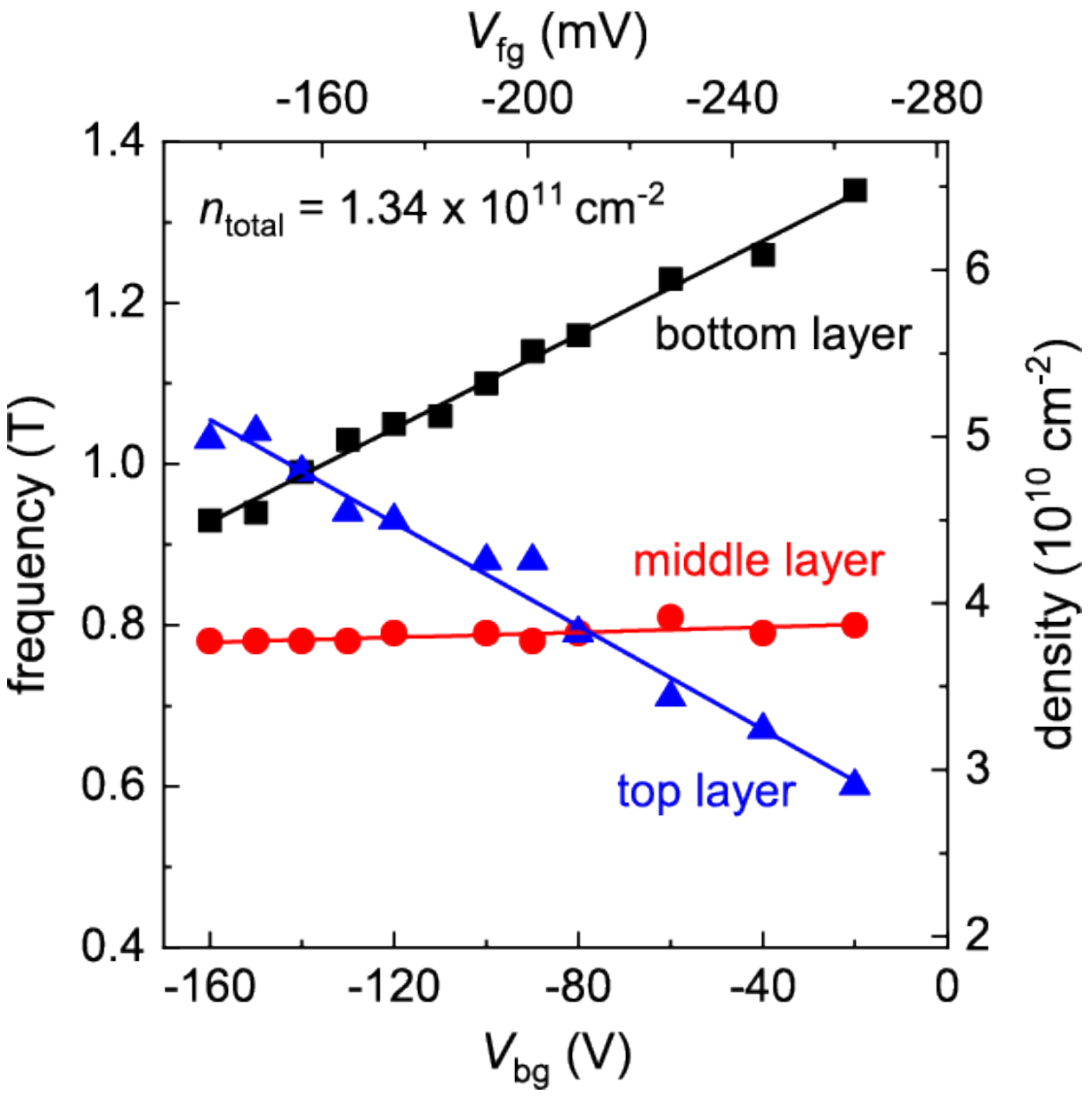


Fig. S2. FFT peak positions as a function of back-gate voltage $V_{bg}$ (bottom axis) and front-gate voltage $V_{fg}$ (top axis). The right axis converts frequency to layer density using $n = 2fe/h$. The solid lines are linear fits to the data points.

red, and blue dashed lines in Fig. S1 are guides to the eyes and track the FFT peak positions for the bottom, middle, and top layers, respectively. At zero $\Delta n$ (when $n_{top} \simeq n_{bottom}$), we observe only two major FFT peaks because the peaks for the top and bottom layers overlap, leading to one stronger peak. Similarly, at $\Delta n/n_{total} = +13\%$ (when $n_{top} \simeq n_{middle}$), only two major peaks are resolved in the FFT spectrum.

In Fig. S2, we plot the FFT peak positions obtained from Fig. S1 as a function of back-gate voltage $V_{bg}$ (bottom axis) and front-gate voltage $V_{fg}$ (top axis). The right axis converts frequency to layer density using $n = 2fe/h$, where $e$ is the electron charge and $h$ is the Planck constant. As expected, increasing $V_{bg}$ (equivalently decreasing $V_{fg}$) increases the bottom-layer density and decreases the top-layer density, while the middle-layer density remains approximately unchanged. The solid lines are linear fits to the data points. The intersection of the black and blue lines determines the gate voltages ($V_{bg} = -140$ V and $V_{fg} = -156$ mV) corresponding to the symmetric charge distribution.

## 2. ADDITIONAL TRANSPORT DATA

### 2.1. Hall resistance measurements

Figure S3(a) shows Hall resistance $R_{xy}$ vs $B$ traces for our trilayer electron system at $n_{\text{total}} = 1.34 \times 10^{11}$ cm$^{-2}$ with symmetric charge distribution, taken for opposite polarities of magnetic field ($B^{+}$ and $B^{-}$) at $T \simeq 30$ mK. Note that while the $R_{xy}^{B+}$ and $-R_{xy}^{B-}$ traces overlap for $B < 4$ T, they significantly deviate from the linear trend, and in opposite directions for $B > 4$ T where the sample becomes highly resistive; see Fig. S3(c) for longitudinal

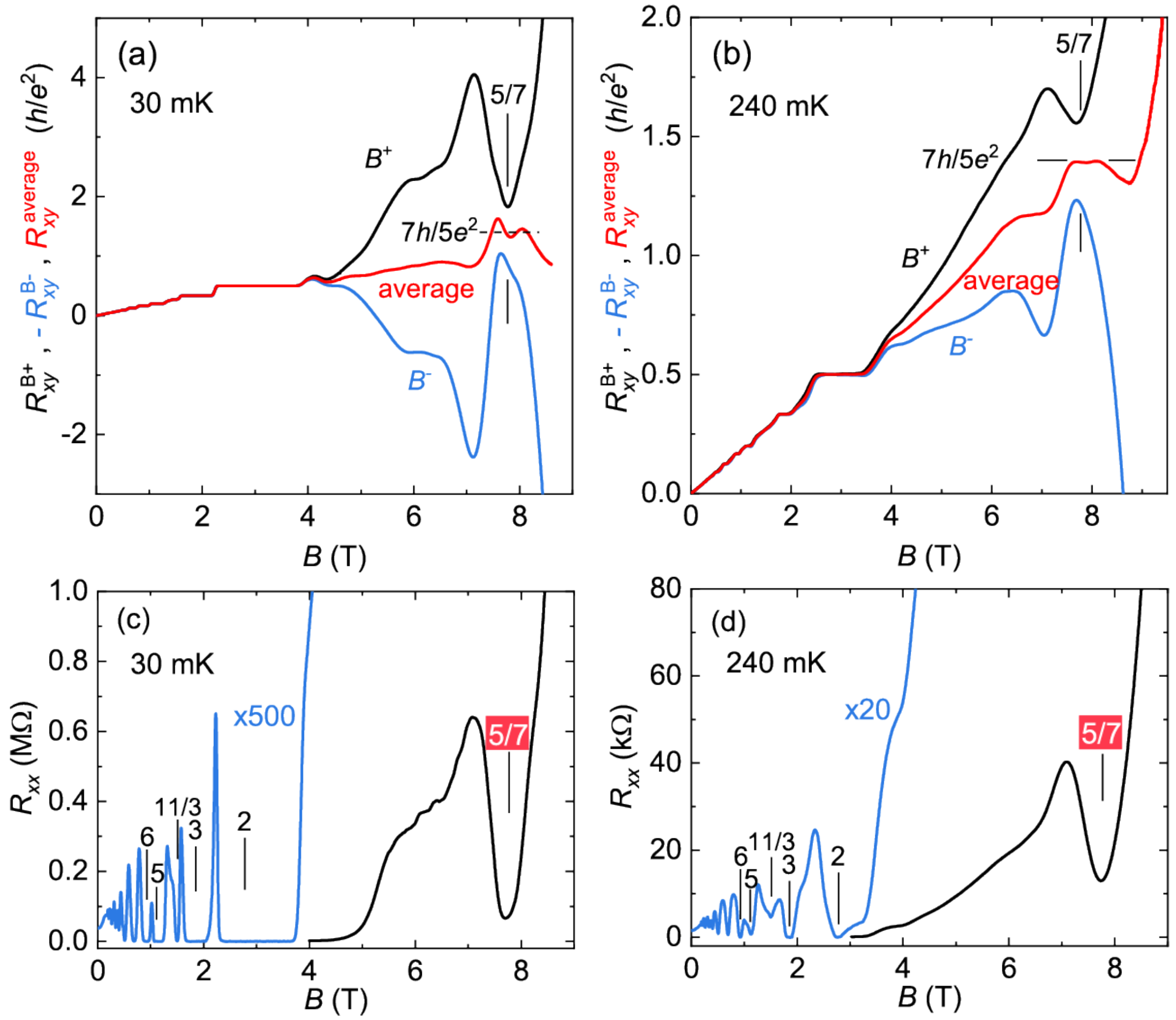


Fig. S3. (a, b) $R_{xy}^{B+}$, $-R_{xy}^{B-}$, and $R_{xy}^{\text{average}}$ vs $B$ traces for $n_{\text{total}} = 1.34 \times 10^{11}$ cm$^{-2}$, taken at (a) $T \simeq 30$ mK, and (b) $T \simeq 240$ mK. The black and blue traces are measured for opposite polarities of magnetic field ($B^{+}$ and $B^{-}$), and the red trace is their average. (c, d) $R_{xx}$ vs $B$ traces taken at (c) $T \simeq 30$ mK, and (d) $T \simeq 240$ mK.

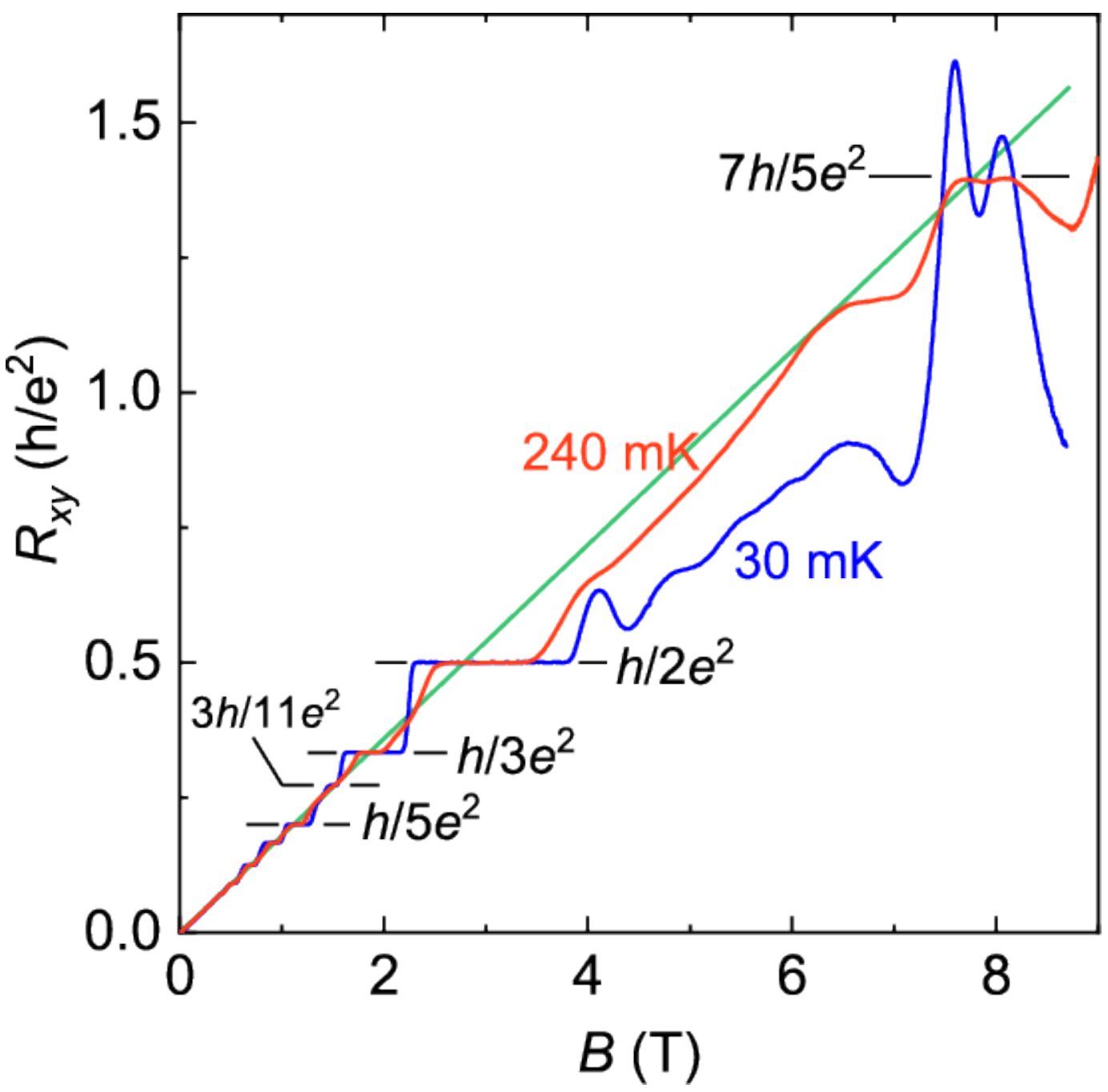


Fig. S4. $R_{xy}$ vs $B$ traces obtained from averaging $R_{xy}^{B+}$ and $-R_{xy}^{B-}$ data at $T \simeq 30$ mK (blue) and 240 mK (red), for $n_{\text{total}} = 1.34 \times 10^{11}$ cm$^{-2}$. The green line shows the expected, classical linear $R_{xy}$ based on the slope of the low-$B$ $R_{xy}$ trace.

resistance $R_{xx}$ vs $B$ trace measured at $T \simeq 30$ mK (The $R_{xx}$ traces for $B^+$ and $B^-$ are symmetric to within 7%.). Such large $R_{xx}$ inevitably induces non-uniformities in the current distribution in the sample and can also cause severe $R_{xx}$ mixing to the Hall traces. The effect of $R_{xx}$ mixing can be minimized by averaging $R_{xy}^{B+}$ and $-R_{xy}^{B-}$ taken for opposite polarities of $B$ [1, 2]. We did such averaging, and find that $R_{xy}^{\text{average}}$ at $\nu = 5/7$ approaches $7h/5e^2$ to within 5%, although a quantized plateau is not seen. This is not surprising because when the sample is highly insulating ($R_{xx} >> h/e^2$), both the current distribution non-uniformities and minor residual $R_{xx}$ mixing can cause problems in Hall measurements.

To further reduce the effect of $R_{xx}$ mixing and non-uniformities in the current distribution induced by the large $R_{xx}$, we also measured $R_{xy}$ for opposite polarities of $B$ at a higher temperature of 240 mK [Fig. S3(b)]. At this temperature, $R_{xx}$ on the flanks of $\nu = 5/7$ is only $\sim 50$ k$\Omega$ ($\sim h/e^2$), more than an order of magnitude smaller than the 30 mK data, and a deep $R_{xx}$ minimum is seen at $\nu = 5/7$; see Fig. S3(d) for the $R_{xx}$ vs $B$ trace measured at 240 mK (The $R_{xx}$ traces for $B^+$ and $B^-$ are symmetric to within 3%.). After averaging,

we observe an $R_{xy}^{\text{average}}$ plateau at $\nu = 5/7$, quantized at $7h/5e^2$ to within 0.8%, establishing a fractional quantum Hall state (FQHS) at $\nu = 5/7$. In Fig. S4, we present the $R_{xy}$ traces after averaging at 30 mK and 240 mK.

### 2.2. $R_{xx}$ and $R_{xy}$ data for the trilayer electron system for different $n_{\text{total}}$ when charge distribution is symmetric

Figure S5 shows $R_{xx}$ and $R_{xy}$ vs $B$ traces measured at $T \simeq 30$ mK, for $n_{\text{total}} = 1.69 \times 10^{11}$ cm$^{-2}$ when the charge distribution is symmetric. Here, the $R_{xy}$ trace is measured with $B^+$ polarity. No signature of $R_{xx}$ mixing is seen because $R_{xx}$ remains relatively small ($<< h/e^2$). As we show in Figs. 3(a) and 3(f) in the main text, at low $B$, the top and bottom layer densities are approximately twice the middle layer density; also see self-consistently calculated charge distribution in Fig. S5 inset. We observe fully developed integer and fractional QHSs at total filling factor $\nu_{\text{total}} = 5$, 2, 4/3, and 2/3, evinced by vanishing $R_{xx}$ and Hall plateaus quantized at $R_{xy} = h/(\nu_{\text{total}} e^2)$. Weaker FQHSs are seen at $\nu_{\text{total}} =$

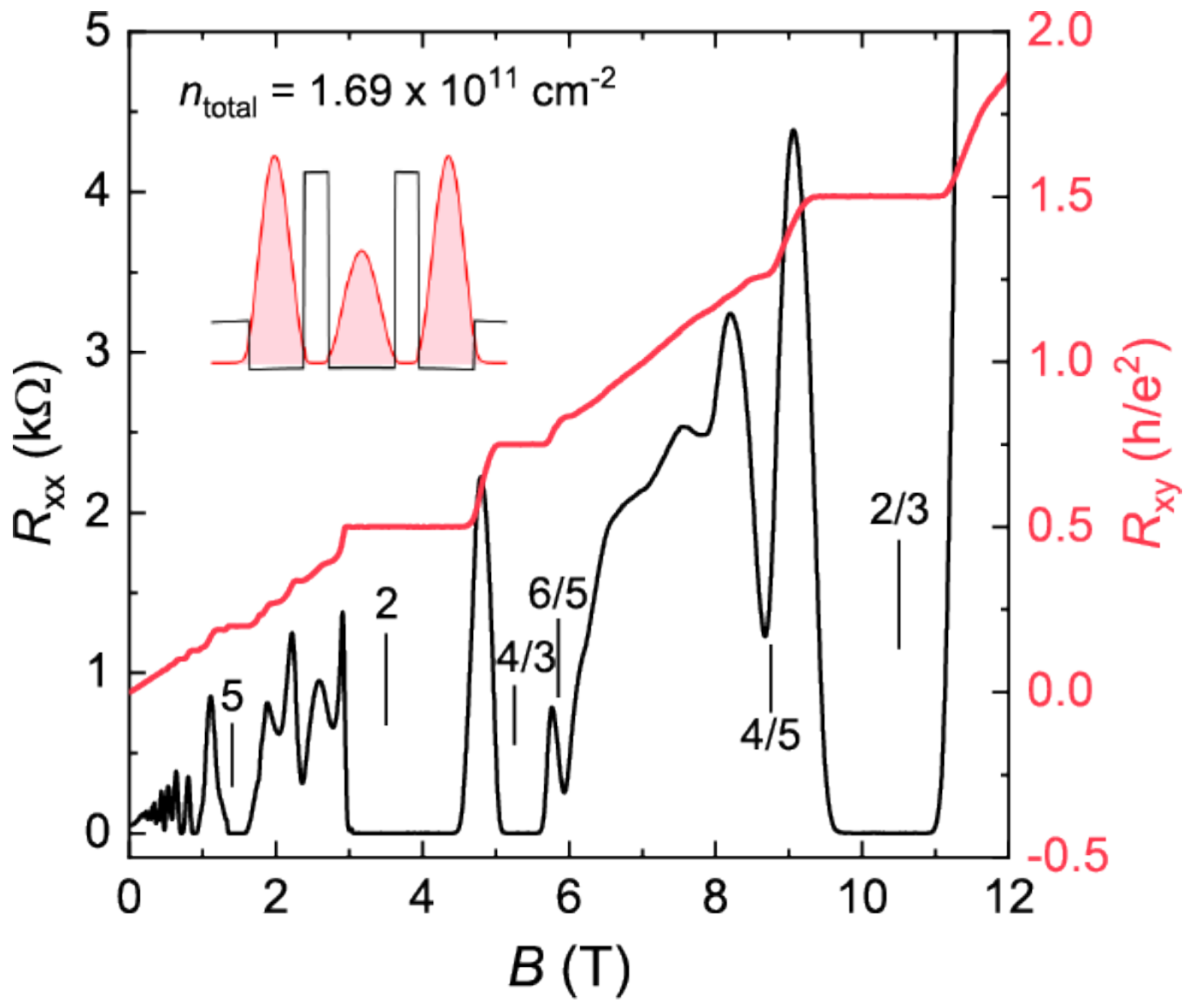


Fig. S5. $R_{xx}$ and $R_{xy}$ vs $B$ traces measured at $T \simeq 30$ mK, for $n_{\text{total}} = 1.69 \times 10^{11}$ cm$^{-2}$. Here, the $R_{xy}$ trace is measured with $B^+$ polarity. No signature of $R_{xx}$ mixing is seen because $R_{xx}$ remains relatively small ($<< h/e^2$). Inset shows the self-consistently calculated charge distribution at zero $B$.

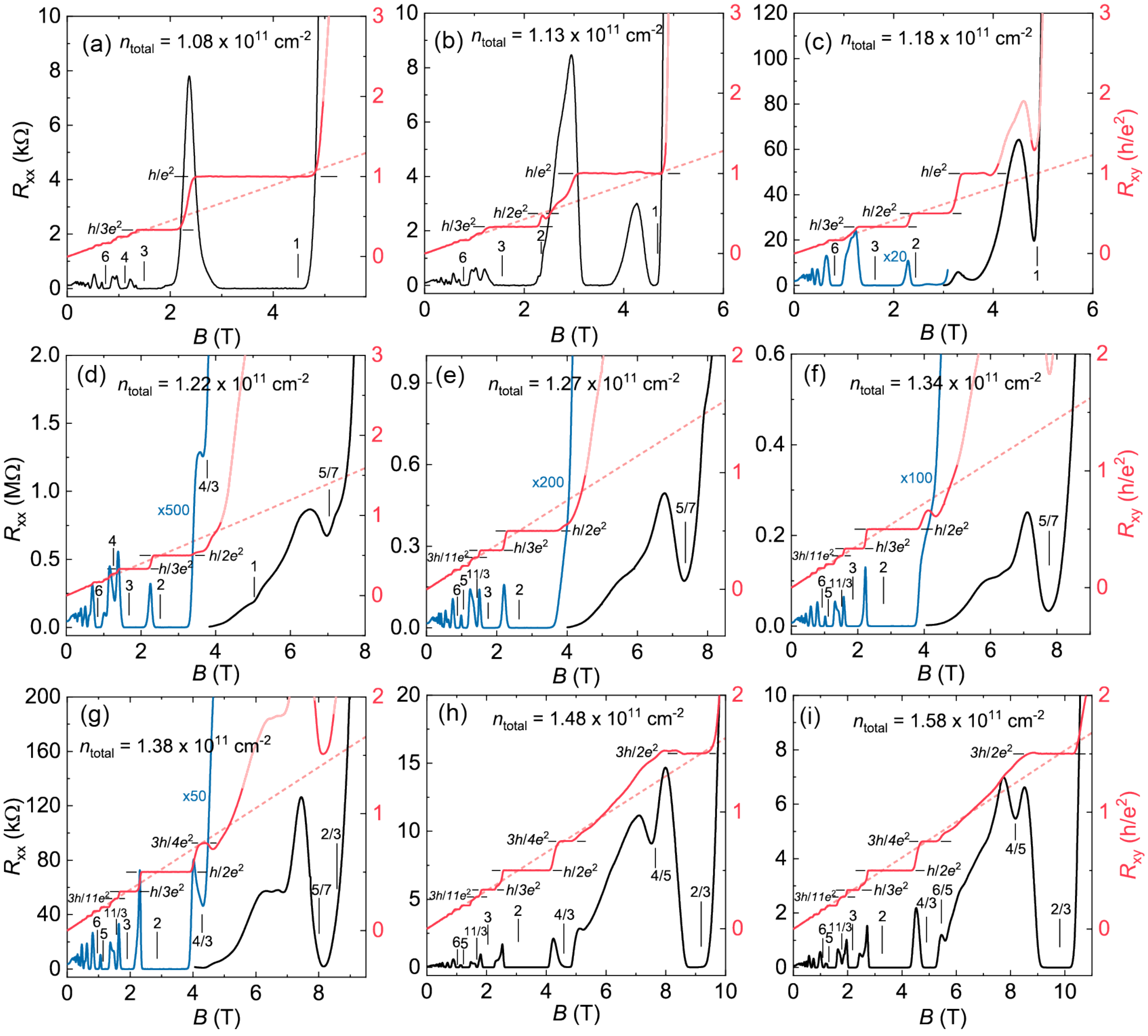


Fig. S6. $R_{xx}$ and $R_{xy}$ vs $B$ traces measured at $T \simeq 30$ mK, for different $n_{\text{total}}$. The charge distribution is kept symmetric ($n_{\text{top}} = n_{\text{bottom}}$). Here, all the $R_{xy}$ traces are measured with $B^+$ polarity. The $R_{xy}$ trace is shown in light red for field ranges when $R_{xx} > h/e^2$ and significantly affects the $R_{xy}$ measurements. The dotted red line shows the expected, classical linear $R_{xy}$ based on the slope of the low-$B$ $R_{xy}$ trace.

6/5 and 4/5. The strong integer QHS observed at $\nu_{\text{total}} = 5$ is consistent with the layer density ratio 2 : 1 : 2 which can support integer QHS in three independent layers with $\nu_{\text{top}} = \nu_{\text{bottom}} = 2$ and $\nu_{\text{middle}} = 1$. At $\nu_{\text{total}} < 2$ ($B \gtrsim 3$ T), the $R_{xx}$ and $R_{xy}$ traces behave similarly to those of a bilayer system. The FQHSs are all observed at *even-numerator* $\nu_{\text{total}}$=

$4/3 = (2/3+2/3)$, $6/5 = (3/5+3/5)$, $4/5 = (2/5+2/5)$, and $2/3 = (1/3+1/3)$, suggesting that all the electrons are in the top and bottom layers [3, 4].

$R_{xx}$ and $R_{xy}$ traces for $n_{\text{total}}$ ranging from $1.08\times10^{11}$ cm$^{-2}$ to $1.58\times10^{11}$ cm$^{-2}$ are shown in Fig. S6. Here, all the traces are measured with $B^{+}$ polarity. At $n_{\text{total}} = 1.22 \times 10^{11}$ cm$^{-2}$, $1.27 \times 10^{11}$ cm$^{-2}$, and $1.34 \times 10^{11}$ cm$^{-2}$, $R_{xx}$ traces exhibit a minimum at $\nu = 5/7$ [Figs. S6(d-f)]. However, $R_{xy}$ near $\nu = 5/7$ significantly deviates from the classical Hall value because the very large $R_{xx}$ induces non-uniformity in the current distribution and severe $R_{xx}$ mixing into the $R_{xy}$ measurements. As we discussed in Sec. 2.1, reliable $R_{xy}$ data near $\nu = 5/7$ are obtained for $n_{\text{total}} = 1.34 \times 10^{11}$ cm$^{-2}$ when the deepest $\nu = 5/7$ $R_{xx}$ minimum is observed, by averaging $R_{xy}^{B+}$ and $-R_{xy}^{B-}$ data at 240 mK; see Fig. S3.

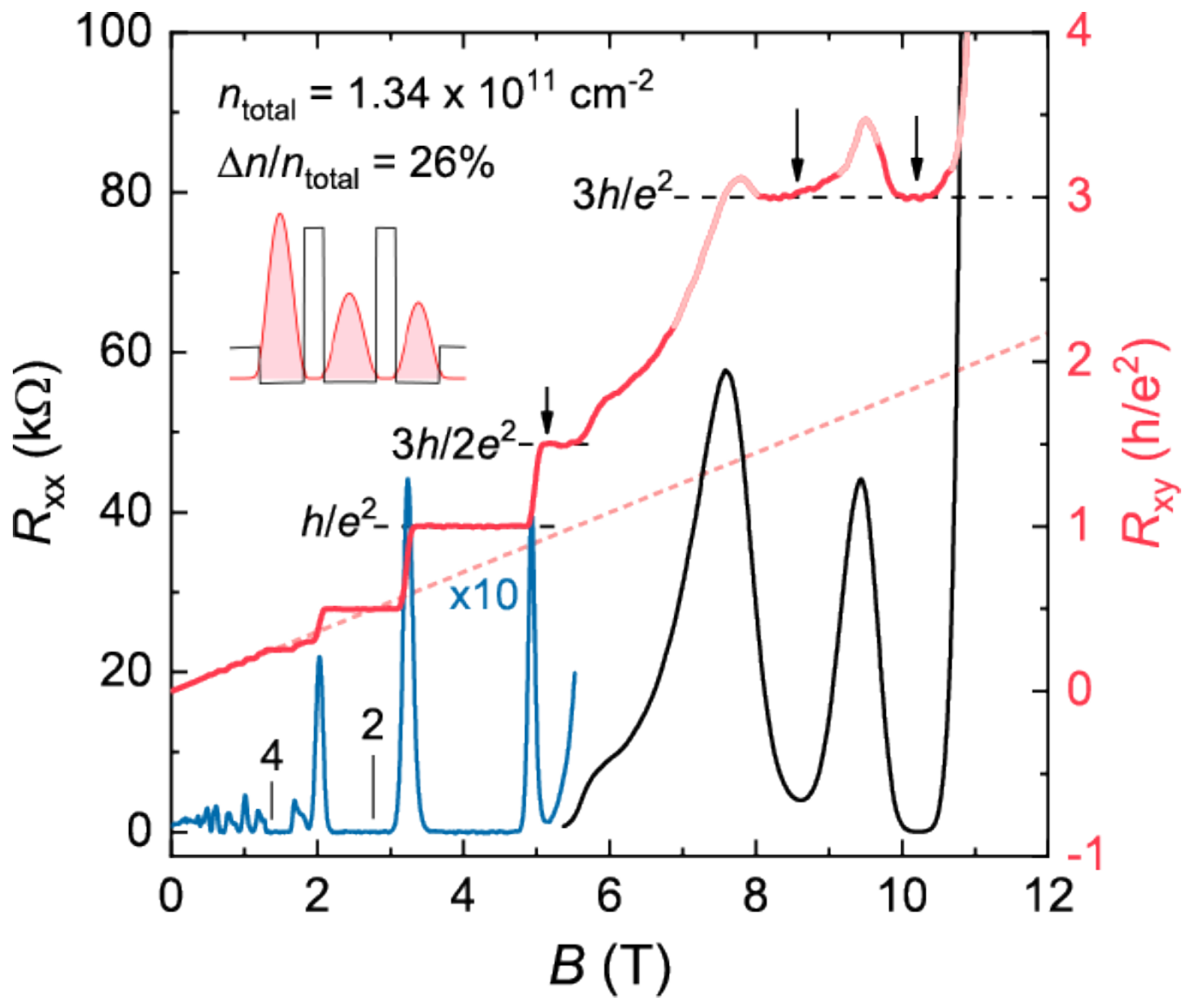


Fig. S7. $R_{xx}$ and $R_{xy}$ vs $B$ traces measured at $T \simeq 30$ mK, for $n_{\text{total}} = 1.34 \times 10^{11}$ cm$^{-2}$ and $\Delta n/n_{\text{total}} = 26\%$. Here, both $R_{xx}$ and $R_{xy}$ trace are measured with $B^{+}$ polarity. The $R_{xy}$ trace is shown in light red for field ranges when $R_{xx} > h/e^2$ and significantly affects the $R_{xy}$ measurements. The dotted red line shows the expected, classical linear $R_{xy}$ based on the slope of the low-$B$ $R_{xy}$ trace. Inset shows the self-consistently calculated charge distribution at zero $B$.

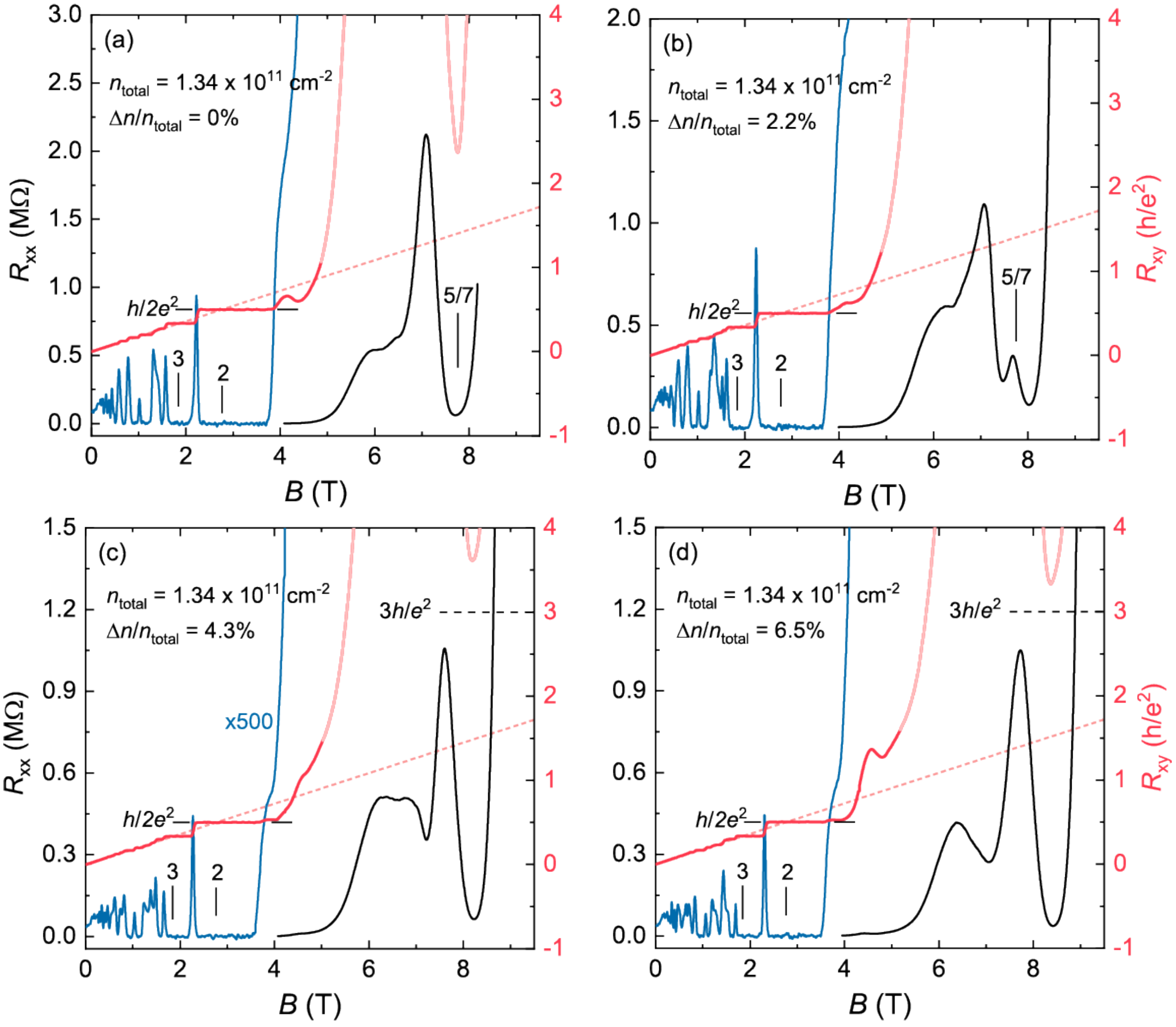


Fig. S8. $R_{xx}$ and $R_{xy}$ vs $B$ traces measured at $T \simeq 30$ mK, for $n_{\text{total}} = 1.34 \times 10^{11}$ cm$^{-2}$, and (a) $\Delta n/n_{\text{total}} = 0\%$, (b) $\Delta n/n_{\text{total}} = 2.2\%$, (c) $\Delta n/n_{\text{total}} = 4.3\%$, and (d) $\Delta n/n_{\text{total}} = 6.5\%$. All $R_{xx}$ and $R_{xy}$ traces are measured with $B^+$ polarity. The $R_{xy}$ trace is shown in light red for field ranges when $R_{xx} > h/e^2$ and significantly affects the $R_{xy}$ measurements. The dotted red lines show the expected, classical linear $R_{xy}$ based on the slope of the low-$B$ $R_{xy}$ trace.

### 2.3. Additional imbalance data for $n_{\text{total}} = 1.34 \times 10^{11}$ cm$^{-2}$

In this section, we present $R_{xx}$ and $R_{xy}$ traces for $n_{\text{total}} = 1.34 \times 10^{11}$ cm$^{-2}$, with different $\Delta n/n_{\text{total}}$. Here, all the traces are measured with $B^+$ polarity. Figure S7 presents data for $\Delta n/n_{\text{total}} = 26\%$. The red dotted line is the classical Hall line for $n_{\text{total}} = 1.34 \times 10^{11}$ cm$^{-2}$.

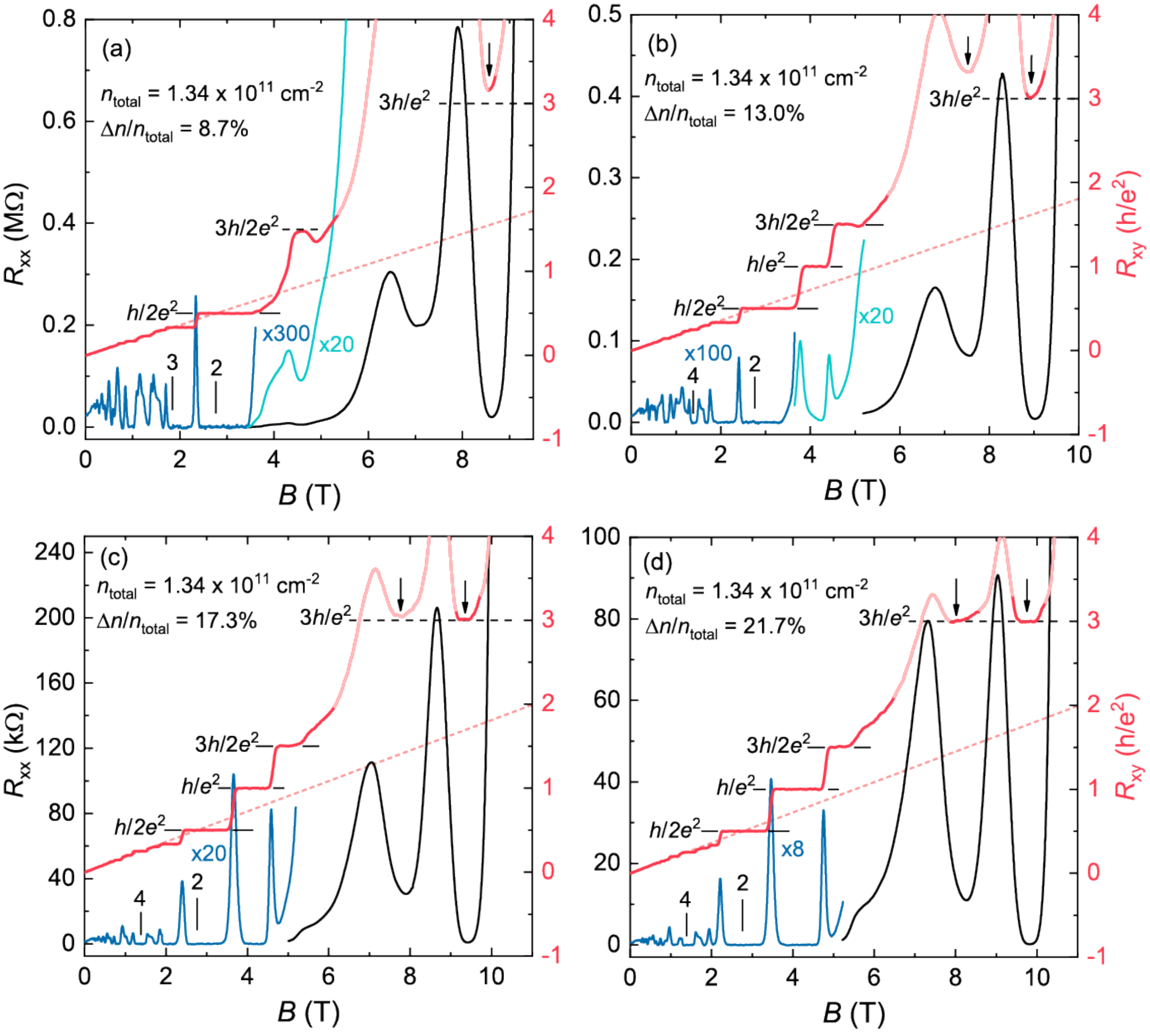


Fig. S9. $R_{xx}$ and $R_{xy}$ vs $B$ traces measured at $T \simeq 30$ mK, for $n_{\text{total}} = 1.34 \times 10^{11}$ cm$^{-2}$, and (a) $\Delta n/n_{\text{total}} = 8.7\%$, (b) $\Delta n/n_{\text{total}} = 13.0\%$, (c) $\Delta n/n_{\text{total}} = 17.3\%$, and (d) $\Delta n/n_{\text{total}} = 21.7\%$. All $R_{xx}$ and $R_{xy}$ traces are measured with $B^+$ polarity. The $R_{xy}$ trace is shown in light red for field ranges when $R_{xx} > h/e^2$ and significantly affects the $R_{xy}$ measurements. The dotted red lines show the expected, classical linear $R_{xy}$ based on the slope of the low-$B$ $R_{xy}$ trace.

For $B < 3$ T ($\nu_{\text{total}} \gtrsim 2$), the $R_{xy}$ vs $B$ trace largely follows the classical Hall line with quantized Hall plateaus appearing for integer QHSs, e.g., at $\nu_{\text{total}} = 4$ and 2. For $B > 3$ T ($\nu_{\text{total}} \lesssim 2$), the $R_{xy}$ vs $B$ trace lies significantly above the classical Hall line. At $B \simeq 4$ T, we observe a vanishing $R_{xx}$ accompanied by a Hall plateau quantized at $h/e^2$. At $B = 5.1$

T, $R_{xx}$ exhibits a sharp minimum while $R_{xy}$ shows a plateau at $3h/2e^2$. At higher $B$, we observe two $R_{xx}$ minima at $B = 8.6$ T and 10.2 T, where $R_{xy}$ exhibits quantized plateaus at $3h/e^2$.

Figures S8 and S9 show the evolution of the $R_{xx}$ and $R_{xy}$ traces as a function of increasing $\Delta n/n_{\text{total}}$. In Fig. S8, we present this evolution as $\Delta n/n_{\text{total}}$ increases from 0% to 6.5%. The $R_{xx}$ minimum at $\nu = 5/7$ observed at $\Delta n/n_{\text{total}} = 0$ splits into two minima when a small $\Delta n/n_{\text{total}}$ (2.2%) is introduced [Figs. S8(a,b)]. These two minima move away from each other with increasing $\Delta n/n_{\text{total}}$ [Figs. S8(b–d)]. The $R_{xy}$ traces near $\nu = 5/7$ are strongly distorted by the very large background $R_{xx}$ ($\gg h/e^2$) and therefore do not provide a reliable indication of the Hall resistance in this regime.

Figures S9(a–d) show the evolution of the $R_{xx}$ and $R_{xy}$ traces as $\Delta n/n_{\text{total}}$ increases from 8.7% to 21.7%. With increasing $\Delta n/n_{\text{total}}$, the two high-$B$ $R_{xx}$ minima (marked by black arrows) become deeper, and the corresponding $R_{xy}$ values approach quantization at $3h/e^2$. Meanwhile, the $R_{xx}$ background near these minima decreases significantly. An $R_{xy}$ plateau at $3h/2e^2$ begins to develop at $\Delta n/n_{\text{total}} = 8.7\%$ and becomes fully quantized for $\Delta n/n_{\text{total}} \gtrsim 13.0\%$. Its position shifts slightly from $B \simeq 4.6$ T at $\Delta n/n_{\text{total}} = 8.7\%$ to a higher magnetic field ($\simeq 5.1$ T) at $\Delta n/n_{\text{total}} = 26\%$. An $R_{xy}$ plateau at $h/e^2$ appears at $B \simeq 4$ T when $\Delta n/n_{\text{total}} \gtrsim 13.0\%$, and its width increases with increasing $\Delta n/n_{\text{total}}$.

---



\end{document}